\documentclass[10pt, conference, letterpaper]{IEEEtran}
\usepackage{amsmath,amssymb,amsfonts}
\usepackage{algorithmic}
\usepackage{graphicx}
\usepackage{textcomp}
\usepackage{xcolor}

\usepackage{twemojis}

\def\BibTeX{{\rm B\kern-.05em{\sc i\kern-.025em b}\kern-.08em
    T\kern-.1667em\lower.7ex\hbox{E}\kern-.125emX}}

\usepackage{xspace}
\usepackage{hyperref}
\usepackage{cleveref}

\usepackage[sorting=none,%
                giveninits=true,%
                maxbibnames=10,%
                minbibnames=1,%
                style=ieee,%
                natbib=true,%
                backend=biber,
                sortcites=true,%
                mincrossrefs=1000]{biblatex}
\AtEveryBibitem{
    \clearfield{editor}
    \clearfield{pages}
    \ifentrytype{book}{
    }{
    \ifentrytype{inbook}{}{
        \clearlist{publisher}
        \clearname{editor}
        \clearfield{isbn}
        \clearfield{issn}
    }}
    \ifentrytype{inproceedings}{
        \clearlist{organization}
        \clearfield{series}
        \clearfield{volume}
        \clearfield{edition}
        \clearfield{doi}
        \clearfield{isbn}
        \clearfield{url}
        \clearfield{pages}
        \clearfield{month}
    }{}
    \clearlist{location}
    \clearlist{address}

    \clearfield{note}
    \clearfield{key}
    \clearfield{doi}
}

\AtBeginBibliography{\footnotesize}

\newcommand{\ie}{i.e., \@}
\newcommand{\eg}{e.g., \@}

\newcommand{\parx}[1]{\noindent\textbf{#1}\xspace}
\newcommand{\takeaway}[1]{\emph{\textbf{To summarize: }{#1}}}

\newcommand{\citehijackstudies}{sermpezis2018artemis,qin2022themis,imaiposter,testart2019profiling,oliver2022stop,biersack2012visual,sermpezis2018survey,cho2019bgp}
\newcommand{\citemoasstudies}{zhao2001analysis,chin2007characteristics,bornhauser2011automatic,jacquemart2014longitudinal,schlamp2015investigating}

\IEEEoverridecommandlockouts

\IEEEpubid{\makebox[\columnwidth]{978-3-903176-58-4~\copyright 2023 IFIP \hfill}\hspace{\columnsep}\makebox[\columnwidth]{ }}

\begin{document}

\title{Live Long and Prosper: \\Analyzing Long-Lived MOAS Prefixes in BGP}

\author{\IEEEauthorblockN{Khwaja Zubair Sediqi}
\IEEEauthorblockA{\textit{Max Planck Institute for Informatics} \\
zsediqi@mpi-inf.mpg.de
}
\and
\IEEEauthorblockN{Anja Feldmann}
\IEEEauthorblockA{\textit{Max Planck Institute for Informatics} \\
anja@mpi-inf.mpg.de
}
\and
\IEEEauthorblockN{Oliver Gasser}
\IEEEauthorblockA{\textit{Max Planck Institute for Informatics} \\
oliver.gasser@mpi-inf.mpg.de
}
}

\maketitle


\begin{abstract}
BGP exchanges reachability information in the form of prefixes, which are usually originated by a single Autonomous System (AS).
If multiple ASes originate the same prefix, this is referred to as a Multiple Origin ASes (MOAS) prefix.
One reason for MOAS prefixes are BGP prefix hijacks, which are mostly short-lived and have been studied extensively in the past years.
In contrast to short-lived MOAS, long-lived MOAS have remained largely understudied.

In this paper, we focus on long-lived MOAS prefixes and perform an in-depth study over six years.
We identify around 24k long-lived MOAS prefixes in IPv4 and 1.4k in IPv6 being announced in January 2023.
By analyzing the RPKI status we find that more than 40\% of MOAS prefixes have all origins registered correctly, with only a minority of MOAS having invalid origins.
Moreover, we find that the most prominent CIDR size of MOAS prefixes is /24 for IPv4 and /48 for IPv6, suggesting their use for fine-grained traffic steering.
We attribute a considerable number of MOAS prefixes to mergers and acquisitions of companies.
Additionally, more than 90\% of MOAS prefixes are originated by two origin ASes, with the majority of detected origin AS relations being customer-provider.
Finally, we identify that the majority of MOAS users are IT companies, and just 0.9\% of IPv4 MOAS and 6.3\% of IPv6 MOAS prefixes are used for anycast.
\end{abstract}


\section{Introduction}
\label{sec:introduction}


To exchange reachability information about IP addresses, different Autonomous Systems (ASes) in the Internet use BGP \cite{rfc4271}.
In BGP, each IP address prefix (\ie collection of IP addresses) is usually originated by a single AS \cite{rfc1930}.
There are, however, also cases where multiple ASes originate the \emph{same prefix}, this prefix is in turn called a Multi Origin AS (MOAS) prefix.
Network operators use MOAS prefixes to \eg provide resilience, load balancing, and multi-homing.

In addition to these uses, MOAS prefixes can also be a result of mergers of companies operating two ASes, misconfigurations \cite{goldberg2014taking}, and---most problematically---prefix hijacks.
This last case occurs if an attacker hijacks traffic to a specific prefix by announcing this prefix with its own AS as an origin.
Unfortunately, prefix hijacks are happening relatively frequently \cite{bgpmon,caida_hijacks_observatory}.
Consequently, prefix hijacks have been the focus of numerous studies over the past years \cite{\citehijackstudies}.

\begin{figure}[!t]
  \centering
  \includegraphics[width=\linewidth]{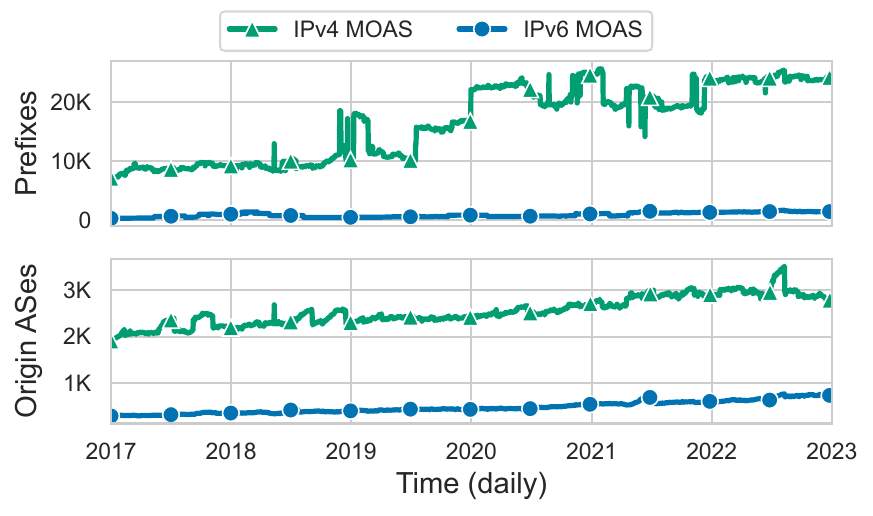}
  \caption{Number of long-lived MOAS prefixes and origin ASes over time.}
\label{int:fig:moas_origins_overtime}
\end{figure}

Surprisingly, the use of long-lived MOAS prefixes---\ie when they are visible for a longer time in BGP---remains understudied.

When analyzing RIB snapshots of all RIPE RIS \cite{ripe2022ris} and Routeviews \cite{routeviews2022data} route collectors, we find that the number of long-lived MOAS prefixes (\ie visible for at least 30 days) is growing over time, as can be seen in \Cref{int:fig:moas_origins_overtime},  both in terms of prefixes and origin ASes.

IPv4 long-lived MOAS prefixes increase from 10k in 2017 to over 24k prefixes at the beginning of 2023, with the number of origin ASes growing by about 50\% in the same time period.
Moreover, the fraction of long-lived MOAS prefixes out of all visible BGP prefixes is increasing as well (see \Cref{subsec:identify-long-lived}).
In this paper, we focus on the analysis of these long-lived MOAS prefixes.
We perform a longitudinal analysis across multiple dimensions to unveil prefix and origin characteristics, visibility, and users of MOAS prefixes.
More specifically, the main contributions of this paper are:

\begin{itemize}
    \item \textbf{Methodology to Detect Long-Lived MOAS Prefixes:}
        We apply a rigorous methodology to detect long-lived MOAS prefixes over a period of six years (see \Cref{sec:method}).
        We use the Kneedle algorithm to discern long-lived from short-lived MOAS prefixes and perform a sensitivity analysis to analyze the influence of route collector artifacts.
    \item \textbf{MOAS Prefixes and Origins:}
        We perform a longitudinal analysis of MOAS prefixes and origin ASes (see \Cref{sec:prefixes-and-origins}).
        We find that more than 40\% of long-lived MOAS prefixes have a valid RPKI status for all origins, hinting that hijacks are not prevalent among them.
        Moreover, we identify that the vast majority of MOAS prefixes are announced by two origin ASes.
        The most commonly used CIDR sizes are /24 for IPv4 and /48 for IPv6, suggesting their use for fine-granular traffic steering.
        Additionally, we find a considerable number of MOAS prefixes linked to mergers of companies.
    \item \textbf{Visibility Analysis:}
        We analyze the visibility of MOAS prefixes across route collectors (see \Cref{sec:visibility}).
        Most MOAS prefixes are visible on hundreds of route collector (RC) peers with one origin AS, whereas the other origin AS is visible only on a handful of RC peers.
    \item \textbf{Users and Usage of MOAS Prefixes:}
        To better understand users and usage of MOAS prefixes we do an in-depth analysis on the AS relationship, their business types, and the prevalence of anycast within MOAS prefixes (see \Cref{sec:users-and-usage-of-moas-prefixes}).
        Our results show that the majority of origin ASes are used by IT companies being in a customer-provider relationship, and only a small fraction of MOAS prefixes are used for anycast.
\end{itemize}

\section{Datasets}
\label{sec:dataset}

To analyze the prevalence and evolution of MOAS prefixes in the Internet, we use the following datasets.


\parx{BGP Datasets:}
To get a good understanding of MOAS prefixes in the Internet, we use route collectors (RCs) from two different route collector projects: RIPE RIS \cite{ripe2022ris} and Routeviews \cite{routeviews2022data}.
We use 59 route collectors throughout our six years study period from January 1, 2017, to January 1, 2023.
During the study period, we analyze daily routing information base (RIB) snapshots of all RCs for the duration of 2098 consecutive days.
To calibrate our methodology (see \Cref{sec:method}) we compile an auxiliary BGP dataset by using three RIBs per day for a period of six months in 2022 from the same RCs.

\parx{AS Relationships Dataset:}
We use CAIDA's AS relationship dataset \cite{caidaASrelationship} from January 1, 2017, to January 1, 2023, to analyze inter-AS relationships of origin ASes associated with MOAS prefixes.
This dataset is released by CAIDA every three months.
The dataset classifies the inter-AS relationships into ``customer-provider or provider-customer'', ``peer to peer'', or ``unknown'' using the method described by Luckie et al. \cite{luckie2013relationships}.

\parx{AS to Organizations Mapping Dataset:}
We use the inferred AS to organization mapping dataset provided by CAIDA \cite{caidaAS2org} for the duration of our study from January 1, 2017, to January 1, 2023.
The dataset is created by querying WHOIS databases from the five Regional Internet Registries (ARIN, LACNIC, RIPE, AFRINIC, and APNIC), and two National Internet Registries (KRNIC and JPNIC) every quarter, and then applying CAIDA's AS to organization technique to map each AS to an organization \cite{caida_AS2org_method}.
Due to the lack of historical data, we are unable to use two recently presented improvements \cite{chen2023improving,arturi20232} to the CAIDA AS to organization dataset.




\parx{ASdb Dataset:}
To gather detailed information about the type of organization using MOAS prefixes, we use the ASdb dataset \cite{ziv2021asdb}.
ASdb provides a mapping of each AS into 17 industry categories and 95 sub-categories using data from established business intelligence databases and machine learning.
We use ASdb to identify AS organization types and business relationships between the origin ASes of MOAS prefixes.

\parx{RPKI Dataset:}
We use archived RPKI records from RIPE NCC \cite{rpkidata} in a monthly granularity from January 2017 to January 2023.
With the RPKI dataset we identify the route origin validation (ROV) state of MOAS prefixes.

\parx{Anycast IP Prefix Dataset:}
Moreover, we use a list of IPv4 and IPv6 prefixes classified as anycast prefixes by the BGP.Tools project \cite{anycasttool}.
The project uses an approach similar to MAnycast$^2$ \cite{sommese2020manycast2} to run periodical anycast detection measurements.
We use the published list of anycast IPv4 and IPv6 prefixes for the most recent date for our study, January 1, 2023, to analyze the prevalence of anycast deployments within detected MOAS prefixes.

\begin{figure*}[!t]
\minipage[t]{0.49\textwidth}
  \centering

  \includegraphics[width=\linewidth]{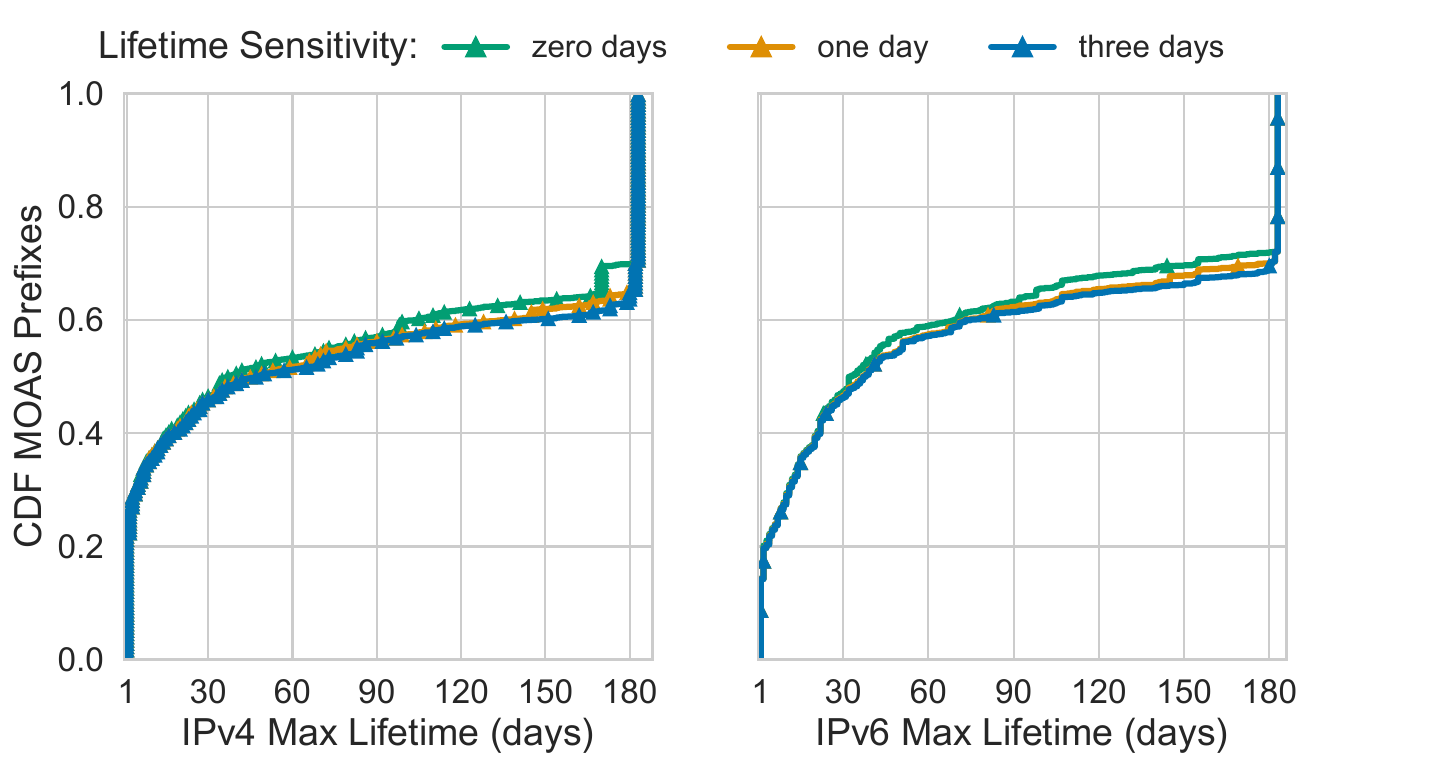}
  \caption{CDF of maximum lifetime with various sensitivity thresholds for IPv4 (left) and IPv6 (right) MOAS prefixes.}
  \label{method:fig:sensitivity_6months}
\endminipage
  \hfill
\minipage[t]{0.49\textwidth}
  \centering
  \includegraphics[width=\linewidth]{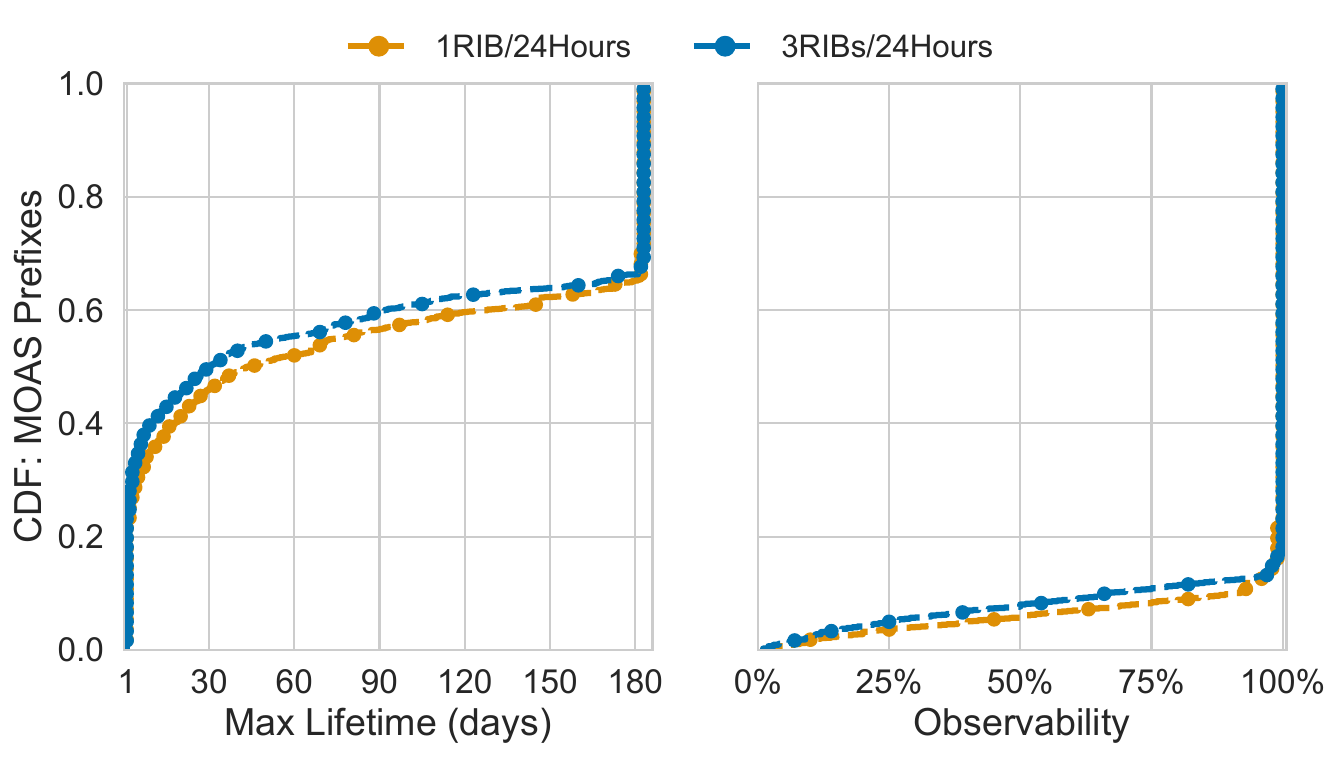}
  \caption{CDF of maximum lifetime (left) and observability (right) of MOAS prefixes for single and three RIBS per day.}
  \label{method:fig:life_span}
\endminipage
\end{figure*}

\section{Methodology}
\label{sec:method}

In this study, we focus on long-lived MOAS prefixes, i.e., prefixes that are announced from multiple origin ASes for a substantial period of time.
In contrast to prefix hijacks \cite{\citehijackstudies}, such MOAS prefixes have received little attention from the research community in the past.


To identify long-lived MOAS prefixes and assess the impact of RC artifacts (\eg missing RIBs, offline RC peers), we apply the following methodology:
%
First, we extract all prefixes from the BGP dataset, apply a series of filters, and select prefixes appearing as MOAS.
Next, we measure the lifetime of MOAS prefixes and perform a sensitivity analysis.
Afterwards, we examine the effect of using more than one RIB snapshot per day in our pipeline.
Finally, we identify and apply a lifetime threshold to distinguish between short-lived and long-lived MOAS prefixes.
All these steps are detailed in the following.

\subsection{Extracting, Filtering, and Selecting MOAS Prefixes}
\label{subsec:extracting-moas}

To develop our methodology, we download and use daily RIB snapshots from all 59 available RCs of the RIPE-RIS \cite{ripe2022ris} and Routeviews \cite{routeviews2022data} projects for a six-months period spanning from June 1, 2022,  to November 30, 2022.
Since we focus on long-lived MOAS prefixes, we do not use BGP UPDATE files, as those provide only additional short-term information.
As a first step, we extract all prefixes from RIB snapshots of RCs and apply a series of filters to remove the following artifacts: private or reserved origin AS numbers \cite{rfc6996}, reserved IP prefixes, prefixes that should not be globally routed, and special purpose IPv4 and IPv6 prefixes \cite{iana-special-ipv4,iana-special-ipv6,rfc5771}.
Furthermore, we remove default route (i.e., \texttt{0.0.0.0/0}) or prefixes with network bits set beyond their CIDR size (i.e., \texttt{1.2.3.0/16}).
As a result of applying filters, we remove around 1\% of all prefixes from the BGP dataset.

Then we check the origin AS of every IP prefix across all RIB snapshots at the exact date and time\footnote{We download available RIBs daily at 08:00:00 UTC from all route collectors. It is important to check the RIB snapshots for the exact date and time across all route collectors, otherwise a prefix could be falsely classified as MOAS appearing with a different origin ASes at two different points in time.
}, select prefixes with more than one origin AS, and consequently mark them as MOAS prefixes for further analysis.

\subsection{MOAS Lifetime Sensitivity Threshold}
\label{subsec:consistency-and-sensitivity}

As BGP data from RC projects can be incomplete, \eg due missing RIBs or RCs temporarily being offline, calculating the lifetime of a MOAS prefixes requires us to be extra cautious.
We use the lifetime, \ie the duration a prefix is seen as a MOAS continuously in RIB snapshots, as a metric to identify and select long-lived MOAS prefixes.
For each MOAS prefix observed in the BGP dataset, we calculate the maximum lifetime over the six months period.
To assess the effect of BGP dataset incompleteness, we perform a sensitivity analysis of the maximum lifetime using different sensitivity thresholds.
%
We investigate how the sensitivity threshold---\ie the maximum number of continuous days a MOAS prefix can be missing from RIB data before its lifetime ends---affects the maximum lifetime of MOAS prefixes.

In \Cref{method:fig:sensitivity_6months}, we show the maximum lifetime for IPv4 and IPv6 MOAS prefixes for sensitivity values of zero, one, and three days.
We find that more than 40\% of IPv4 and around 50\% of IPv6 MOAS prefixes have a maximum lifetime of fewer than 30 days.
On the other end, around 35\% of IPv4 and 30\% of IPv6 MOAS prefixes are visible throughout the entire six months.
Moreover, around 25\% of IPv4 and 20\% of IPv6 MOAS prefixes have a maximum lifetime of at most five days.

Looking at the effect of different sensitivity values, we notice that overall the one and three days sensitivity values slightly increase the lifetime of MOAS prefixes compared to the zero days sensitivity.
This difference is most visible for IPv4 in the interval of 150 to 180 days, where we see the effect of a route collector outage when the RIB snapshot of Sydney RC for time 08:00:00 UTC is missing on June 13, 2022,  but available for the rest of the time.
Considering this and as a precaution to mitigate possible effects of unavailable RCs or RC peers on the lifetime of MOAS prefixes, we use the one day sensitivity threshold in the remainder of the paper.
We do not consider the three days sensitivity threshold, because it does not substantially increase the lifetime of MOAS prefixes beyond the one day threshold, and a gap of three days during which a MOAS prefix is not visible is large enough to possibly lead to incorrect calculations of MOAS prefix lifetimes.

\begin{figure*}[!t]
  \minipage[t]{0.49\textwidth}
  \centering
  \includegraphics[width=\linewidth]{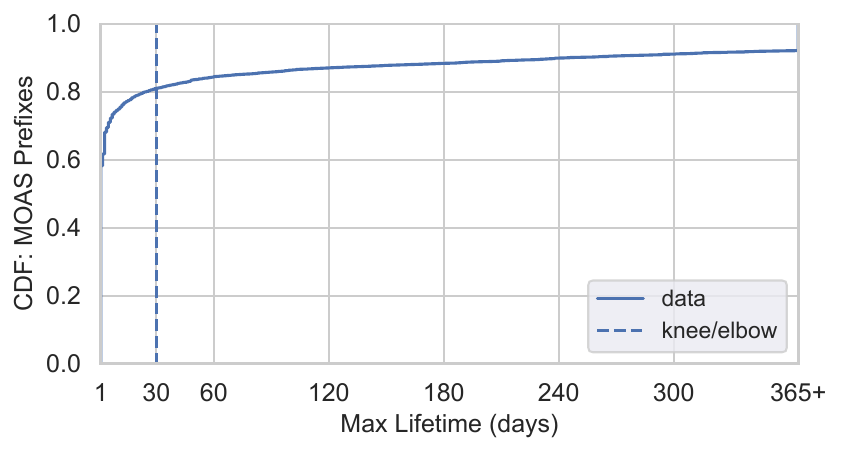}
  \caption{CDF of maximum lifetime of IPv4 and IPv6 MOAS prefixes capped at 365 days, vertical line shows the elbow/knee of the distribution.}
  \label{method:fig:knee}
  \endminipage
\hfill
\minipage[t]{0.49\textwidth}
  \centering
  \includegraphics[width=\linewidth]{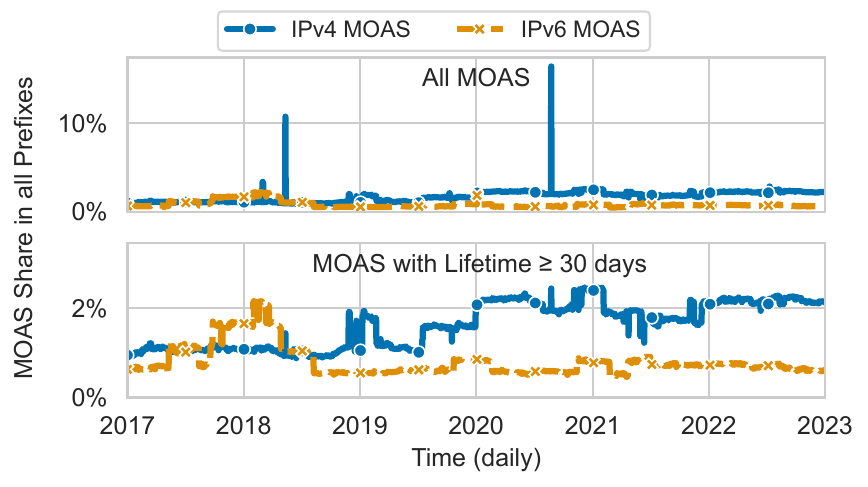}
  \caption{Fraction of MOAS prefixes out of all prefixes for IPv4 and IPv6 over time, all MOAS (top) and long-lived MOAS (bottom).}
\label{int:fig:moas_vs_all_overtime}
\endminipage
\end{figure*}

\subsection{Impact of Using More Than One RIB per Day}
\label{subsec:using-more-than-one-ribs}

After having performed the lifetime analysis with various sensitivity values, we examine if increasing the number of RIB snapshots per day increases the lifetime of MOAS prefixes.
Therefore, we use three RIB snapshots per day (3RIBs/24hours), as this is the maximum number of possible RIB snapshots with the same timestamp across both Routeviews and RIPE RIS RCs.
We study the effect of increasing the number of RIB snapshots considering the following two scenarios.

First, we examine by how much the increased number of RIBs increases the maximum lifetime of MOAS prefixes.
The left subplot of \Cref{method:fig:life_span} illustrates the comparison and effect of using three RIBs compared to one RIB per 24 hours on the maximum lifetime of MOAS prefixes.
As can be seen, increasing the number of RIBs to three per day (blue line) does not substantially increase the maximum lifetime of MOAS prefixes compared to a single RIB per day, orange line.
Furthermore, the number of MOAS prefixes visible throughout the complete six months remains similar.
The higher share of MOAS prefixes with a maximum lifetime of fewer than five days in three RIBs compared to a single one is because we observe many more short-lived MOAS (\eg MOAS appearing only once) when using three RIBs per day.



Second, we check if increasing the number of RIB snapshots increases the observability of MOAS prefixes.
We define the observability of a MOAS prefix as the number of days out of the total days, when a prefix is observed as a MOAS.
Therefore, the observability metric allows us to see how consistently prefixes are visible as MOAS.
As shown in the right subplot of \Cref{method:fig:sensitivity_6months}, for nearly 80\% of MOAS both one and three RIBs have more than 95\% observability.
This high value gives us confidence that using the maximum lifetime of a MOAS prefix is a good approximation of its actual lifetime.
Moreover, with three RIBs we see a higher fraction of MOAS with 100\% observability compared to one RIB.
This is because with three RIBs we observe more MOAS prefixes that are visible only on a single RIB snapshot, which by definition have an observability of 100\%.

To summarize, increasing the number of RIBS from one to three per day does not increase the lifetime or observability of MOAS prefixes, particularly long-lived ones.
Instead, three RIBs per day add more short-lived MOAS, which are not the focus of this study.
Therefore, in the remainder of the paper we use one RIB per day.

\subsection{Identifying Long-Lived MOAS Prefixes}
\label{subsec:identify-long-lived}

Previous studies have shown that prefix hijacks, route leaks, and configuration errors cause short-lived MOAS \cite{\citemoasstudies}.
We strive to analyze long-lived MOAS prefixes in BGP.
Therefore, we develop a technique to identify a threshold to distinguish between short-lived and long-lived MOAS prefixes.

We measure the maximum lifetime of all MOAS prefixes for our complete BGP dataset as observed from January 1, 2017, to January 1, 2023.
Then, we use the mathematical Kneedle algorithm \cite{satopaa2011finding} to determine the ``elbow'' of the maximum lifetime of MOAS prefixes.
The ``elbow'' point identifies the maximum curvature value within the lifetime of all MOAS prefixes and we use it as distinction point between short-lived and long-lived MOAS prefixes.

\Cref{method:fig:knee} shows the maximum lifetime (capped at 365 days) for all MOAS prefixes.
Using the Kneedle algorithm we identify 30 days as the elbow, which we therefore use as a threshold to distinguish between short-lived and long-lived MOAS prefixes.
Generally, we find that more than 80\% of MOAS prefixes are short-lived, whereas less than 20\% have a lifetime of at least 30 days.


Now that we defined the threshold of minimum 30 days lifetime between short-lived and long-lived MOAS prefixes, we examine the overall share of MOAS prefixes in the BGP dataset.
In \Cref{int:fig:moas_vs_all_overtime} we show the fraction of MOAS prefixes among all prefixes for the duration of six years for IPv4 and IPv6, respectively.
The fluctuation in the number of IPv4 MOAS prefixes for the upper subplot is due to Angola Cables (AS37468) creating MOAS conflicts for around 90k prefixes, on May 12, 2018, leading to the 10\% IPv4 MOAS peak.
Moreover, on August 24, 2020, Huge Networks - DDoS Mitigation (AS264409) creates MOAS conflicts for 143k prefixes, resulting in the 15\% peak of IPv4 MOAS prefixes. 

The lower subplot is limited to long-lived MOAS prefixes only, and exhibits less variation.
We see the fraction of MOAS prefixes in IPv4 increasing from around 1\% in 2017 to around 2\% in 2023.
In IPv6, the fraction is relatively stable at just below 1\%, with a peak in February 2018.
This peak is a result of GTT Communications (AS3257, AS5580) and DigitalOcean (AS14061, AS200130, AS202018, AS202109, AS393406, AS62567) each announcing hundreds of IPv6 hyper-specific \cite{sediqi2022hyper} (\ie /49--/128) MOAS prefixes.

In the remainder of the paper we focus exclusively on long-lived MOAS prefixes.
Therefore, we now use the term MOAS prefix interchangeably with long-lived MOAS prefix.

\subsection{Limitations}

While we try to address multiple caveats with our approach, it still has some limitations.
Our observation of MOAS prefixes in the BGP dataset is limited to peer ASes of RCs.
Therefore, our detected MOAS prefixes should be seen as a \emph{lower-bound}, \ie there might be more MOAS prefixes out there which are not visible at route collectors.
As the vast majority---95\% for IPv4 and 91\% for IPv6---of MOAS prefixes are originated by the exact same set of origins throughout the six months, we do not give special treatment to MOAS prefixes with changing AS origins over time.
Despite these limitations, we argue that our approach is suitable and our analysis provides valuable insights into the prevalence, characteristics, and use of MOAS prefixes in the Internet.

\section{Prefixes and Origins}
\label{sec:prefixes-and-origins}

In this section we study MOAS prefixes and origins in detail:
We analyze the RPKI status, CIDR sizes, and origin ASes of MOAS prefixes.

\subsection{RPKI Status of MOAS Prefixes}
\label{subsec:rpki-status-of-moas-prefixes}

As MOAS prefixes by definition are announced by at least two origin ASes, we now investigate whether a MOAS prefix has valid RPKI entries for its origin ASes.
We examine the route origin validation (ROV) state of MOAS prefixes using the RPKI dataset \cite{rpkidata} for all origins---\ie for every prefix-origin (PO) pair---on a monthly basis, starting from January 2017 to January 2023.
We classify the ROV state of MOAS prefixes as follows: (i) ``All Valid Origins'' if all PO pairs have a valid route origin authorization (ROA) record; (ii) ``At Least One Valid Origin'' if at least one PO pair of a MOAS prefix has a valid ROA record; (iii) ``All Invalid Origins'' if all PO pairs conflict with a present ROA record; (iv) ``Not Found'' if we do not observe any ROA for a MOAS prefix.

\begin{figure}[!t]
  \centering
  \includegraphics[width=\linewidth]{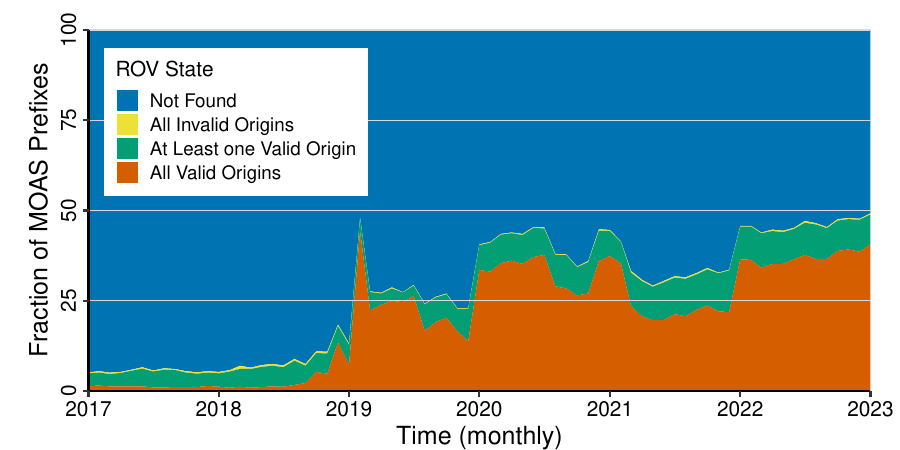}
  \caption{MOAS Route Origin Validation state in the RPKI dataset over time.}
  \label{fig:function:moas_in_rpki}
\end{figure}

In  \Cref{fig:function:moas_in_rpki} we show the RPKI classification of MOAS prefixes over time.
We note that the percentage of MOAS prefixes with ``All Valid Origins'' (shown in orange), has increased from less than 5\% in 2017 to around 40\% in January 2023.
The increase in share of valid ROV states is in line with the RPKI deployment in all BGP prefixes \cite{rpkimonitor}.
It also underlines that a large part of MOAS prefixes are not due to prefix hijacks, route leaks, or router configuration errors, but are in fact intended for use.
We see a peak in the beginning of 2019 for ``All Valid Origins'', which reaches around 45\% and is due to TTNet (AS47331) and Turk Telekomunikasyon (AS9121) announcing nearly 8k IPv4 MOAS prefixes---mostly /24 and /21--/23---as a result of their merger \cite{dailysabah}.
The share of MOAS prefixes with at least one valid origin (shown in green) has slightly increased from 2017 to 2023.
Manually investigating the partially valid cases, we notice that not all origin ASes entered the prefix information properly in the RPKI database.
We find cases where both ASes are registered with same organization name, with one having a valid and the other having an invalid ROV state for the MOAS prefix.
Another possible reason for partially valid MOAS prefixes are router misconfigurations.
The chance of them being hijacked prefixes is relatively small, as hijacks usually have short durations \cite{\citehijackstudies} and our analyzed prefixes are seen for at least 30 days.
MOAS prefixes with all invalid origins (shown in yellow), contribute less than 1\% throughout our measurement period.
The share of MOAS prefixes for which no ROA entry exists (shown in blue) decreases from 90\% in 2017 to around 50\% in 2023.

\takeaway{Around 40\% of MOAS prefixes in January 2023 have all their origins validated in RPKI, with another $\approx$10\% having a partially valid RPKI status.
The case of all origins being invalid occurs for less than 1\% of cases.
This underlines that a large part of long-lived MOAS prefixes is likely intended for actual use and not linked to hijacking, misconfigurations, or route leaks.}

\subsection{CIDR Sizes}

\begin{figure}[!t]
  \centering
  \includegraphics[width=\linewidth]{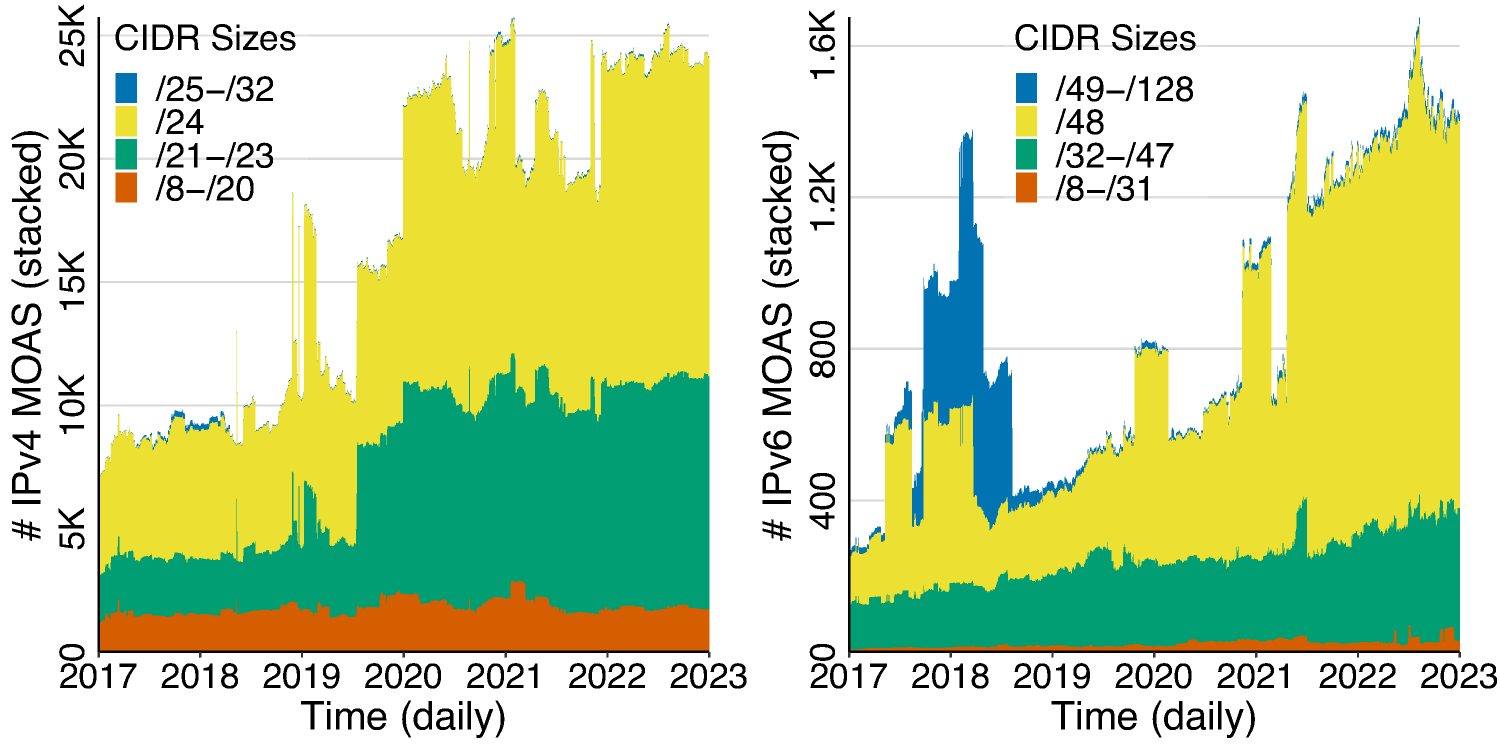}
  \caption{MOAS prefix CIDR sizes over time, for IPv4 (left) and IPv6 (right).}
  \label{fig:long:moas_cidrs}
\end{figure}

Prefixes in BGP are announced at different granularity levels, \ie CIDR sizes.
These CIDR sizes can provide insights into the usage of prefixes.
Therefore, we analyze CIDR sizes of MOAS prefixes as shown in \Cref{fig:long:moas_cidrs} (IPv4 on the left, IPv6 on the right).
For IPv4, we find that MOAS prefixes overall have increased from fewer than 10k in 2017 to more than 24k in 2023.
The /24 CIDR size is most prominent in IPv4, followed by the /21--/23 CIDR size group.
Hyper-specific \cite{sediqi2022hyper} MOAS prefixes, \ie /25--/32, constitute a tiny fraction of MOAS prefixes, as shown in blue.

While the number of MOAS prefixes with a CIDR size of /8--/20 is relatively stable over time, both the /24 and the /21--/23 range increase by more than 50\% after mid-2019.
We investigate the peaks in the data and attribute them to the merger of TTNet and Turk Telekomunikasyon, see \Cref{subsec:rpki-status-of-moas-prefixes}.
These prefixes are visible as a substantial increase in the /21--/23 CIDR size group from August 2019 to January 2023.
In addition, the peaks of /24 MOAS prefixes in July 2019, January 2020, September 2020, and January 2022 are associated with thousands of MOAS prefixes originating from Orange Spain (AS12479, AS12715).
The latter of the two ASes used to belong to Jazztel, which was acquired by Orange in 2014 \cite{orangejazztel}.
Moreover, the acquisition of KPN International (AS286) by GTT (AS3257) in December 2019 \cite{gttkpn} still results in 389 /24 MOAS prefixes in January 2023.

In IPv6 we see an increase from fewer than 400 to more than 1400 MOAS prefixes, as shown in the right plot of \Cref{fig:long:moas_cidrs}.
The number of /48 MOAS shown in yellow, which forms the largest CIDR size in IPv6, increases more than five-fold from 2017 to 2023.
The CIDR size ranges /32--/47 and /8--/31 form the second and third-largest group of IPv6 MOAS prefixes and grow steadily.
We see a large number of IPv6 hyper-specific MOAS prefixes, \ie /49--/128 CIDR sizes shown in blue, from mid-2017 to mid-2018, which are normally not routable in the Internet \cite{sediqi2022hyper}.
These are again linked to GTT and DigitalOcean, see \Cref{subsec:identify-long-lived}.

\takeaway{ASes use mostly fine-granular CIDR sizes---\ie /24 for IPv4 and /48 for IPv6---to announce MOAS prefixes in the Internet, which allows for more fine-grained control of routing policies.
Overall, the distribution of CIDR sizes within MOAS prefixes is similar to non-MOAS prefixes reported by BGP routing table \cite{bgppotaroo}.
Moreover, we find that a considerable number of MOAS prefixes are related to mergers and acquisitions of companies.}


\subsection{Origin ASes}

\begin{figure}[!t]
\centering
\includegraphics[width=\linewidth]{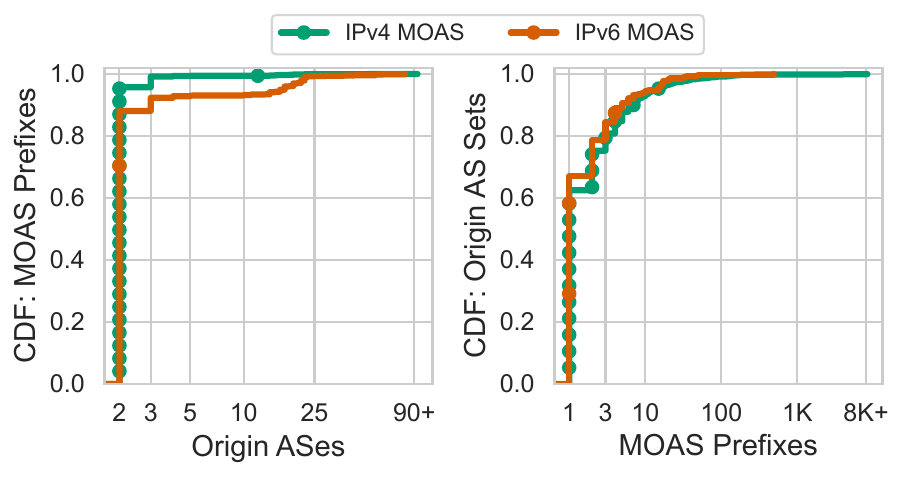}\label{fig:origin_p_prefix}
  \caption{Number of origin ASes per prefix (left) and the number of originated MOAS prefixes per origin AS set (right) for January 1, 2023.}
\label{long:fig:origin_per_prefix}
\end{figure}

Next, we analyze the origin ASes for every MOAS prefix in more detail.
In the left subplot of \Cref{long:fig:origin_per_prefix} we observe that more than 95\% of IPv4 and around 88\% of IPv6 prefixes have only two origin ASes (\ie the minimum required for them to be classified as a MOAS prefix).
Moreover, around 1\% of IPv4 and 8\% of IPv6 MOAS prefixes have more than three origin ASes.
We notice that a few prefixes are originated by more than 80 origin ASes.
We manually investigate these cases and identify Verisign---a DNS provider which also operates two DNS root servers---to have the highest number of origin ASes for one IPv4 and one IPv6 prefix.

Furthermore, we analyze the number of MOAS prefixes originating from the same set of origin ASes, so-called origin AS sets.
In \Cref{long:fig:origin_per_prefix} the right plot illustrates that more than 60\% of origin AS sets announce only a single MOAS prefix and around 90\% announce fewer than ten MOAS prefixes.
In addition to January 1, 2023, we also check one day per month and observe a similar pattern.
Moreover, only a few origin AS sets announce large numbers of MOAS prefixes.
Investigations show that this is again the TTNet and Turk Telekomunikasyon merger case, see \Cref{subsec:rpki-status-of-moas-prefixes}.

\takeaway{The vast majority (95\% for IPv4, 88\% for IPv6) of MOAS prefixes are originated by two origin ASes, with some outlier prefixes being announced by more than 80 origins.
Around 90\% of origin AS sets announce fewer than ten MOAS prefixes.
Therefore, the typical case is that two origin ASes announce a small number of MOAS prefixes.}

%

\section{Visibility}
\label{sec:visibility}

In this section we analyze the visibility of MOAS prefixes across route collector peers and investigate the minimum and maximum visibility per prefix in detail.

\subsection{Visibility Across Route Collector Peers}

\begin{figure}[!t]
    \centering
    \includegraphics[width=\linewidth]{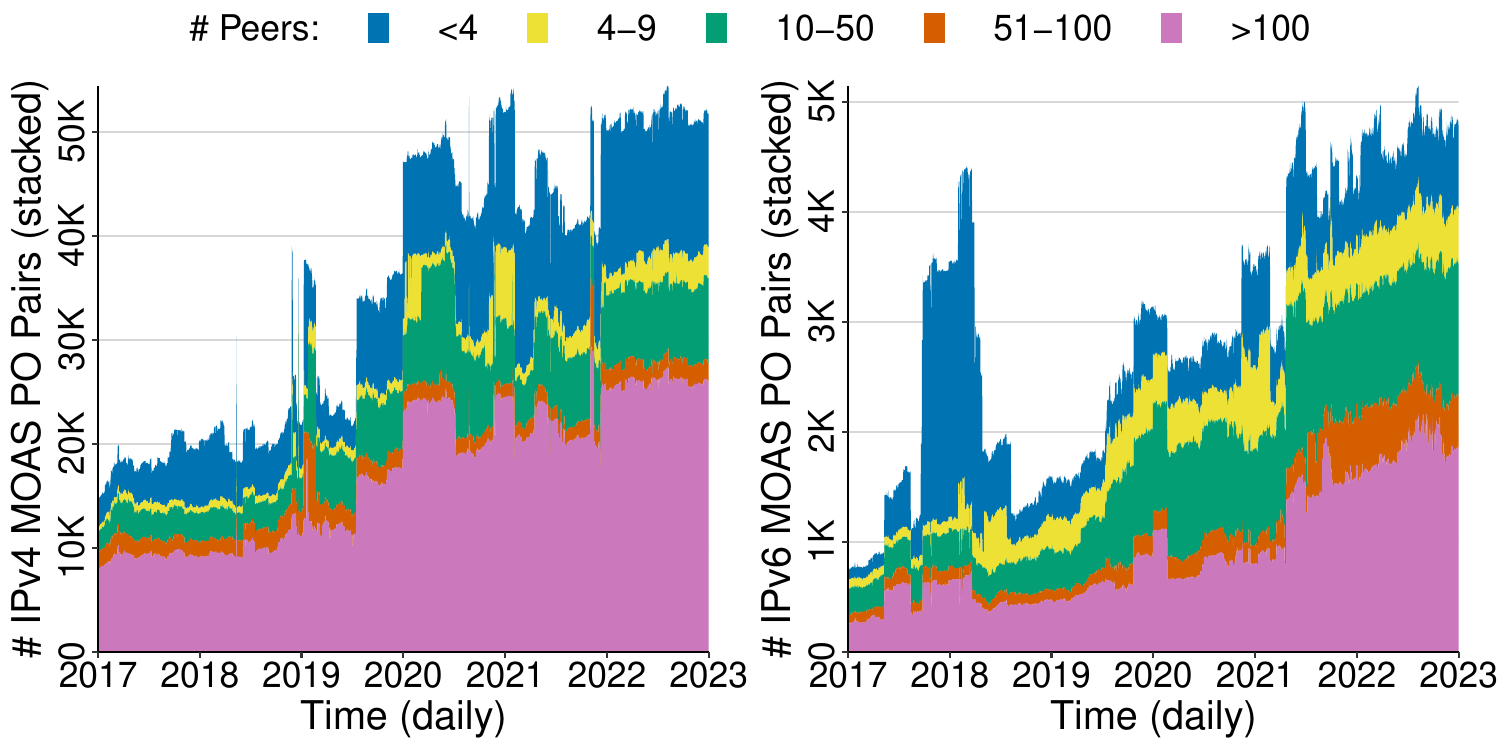}
    \caption{MOAS prefix-origin pair visibility across route collector peers over time, for IPv4 (left) and IPv6 (right).}
    \label{fig:long:po_visibility}
\end{figure}

First, we focus on the visibility of MOAS prefixes in route collector peers.
RC peers are the ASes that feed routing information to the route collector via BGP sessions.
\Cref{fig:long:po_visibility} illustrates the visibility of every MOAS prefix-origin (PO) pair, for IPv4 (left plot) and IPv6 (right plot), across RC peers for the period of the study.
For IPv4, the largest group of MOAS PO pairs (\ie around 50\%), is visible in more than one hundred route collector peers (shown in pink), followed by MOAS PO pairs with a visibility of 3 or fewer peers (shown in blue).
This ``bimodal'' visibility distribution is consistent with overall BGP prefixes \cite{testart2019profiling}.
PO pairs with a visibility between 10 and 50 peers (shown in green) form the third largest group.

In IPv6 we see a similar picture, with MOAS PO pairs with visibility $>$100 being most common, followed by PO pairs with visibility between 10 and 50 peers.
The $<$4 category forms the third-largest group during our study period.
However, from mid-2017 to mid-208, we see a visible spike in IPv6 to around three thousand PO pairs in the $<$4 category.
We investigate this peak manually and find that it is largely due to hyper-specific MOAS prefixes originated by GTT Communications and DigitalOcean (see \Cref{subsec:identify-long-lived}) and /48 MOAS prefixes originated from the Thai eyeball network Realmove Company Limited (AS132061, AS7470).

\takeaway{PO pairs of IPv4 MOAS exhibit a ``bimodal'' RC visibility behavior:
Either they are widely visible in more than a hundred RC peers or very poorly visible in fewer than four RC peer.
For IPv6 we see a similar picture, although semi-visible PO pairs are more common than in IPv4.}


\subsection{Minimum and Maximum Visibility}

\begin{figure}[!t]
  \centering
  \includegraphics[width=\linewidth]{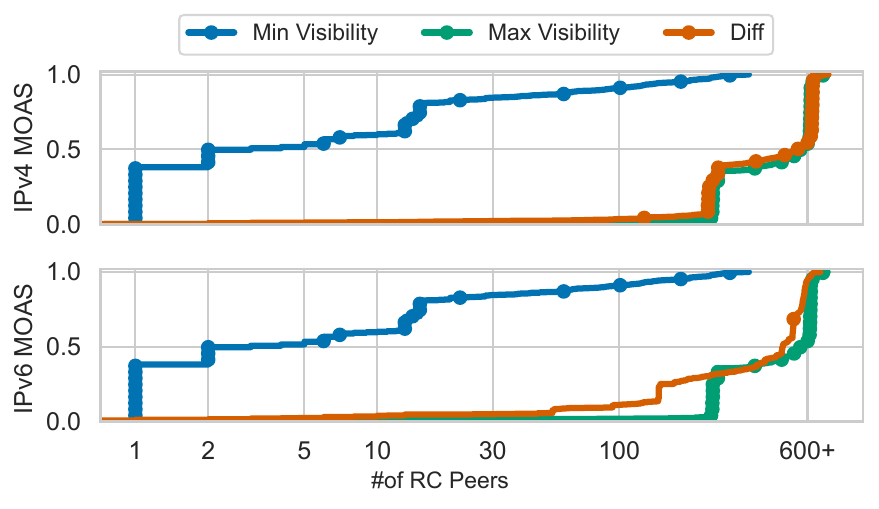}
  \caption{Minimum, maximum, and difference visibility of MOAS prefixes on January 1, 2023 for IPv4 (top) and IPv6 (bottom).}
  \label{long:fig:max_min_visibility}
\end{figure}

To better understand the bimodal visibility behavior of MOAS prefixes, we investigate the minimum and maximum visibility (\ie number of route collector peers) of every MOAS prefix depending on its origin AS.
This allows us to see if a MOAS prefix has a similar visibility for all its origins or if the minimum and maximum are widely spread.
In \Cref{long:fig:max_min_visibility} we show the minimum and maximum visibility of each PO pair for IPv4 (top) and IPv6 (bottom) for a single day (\ie January 1, 2023).
We also check one day per month for the entire period of study, and observe a similar pattern.


For the maximum visibility we observe that for 99\% of all MOAS prefixes---in both IPv4 and IPv6---at least one PO pair is visible by more than 100 RC peers.
Moreover, for more than 50\% of MOAS prefixes, at least one PO pair has a maximum visibility larger than 500 route collector peers.
We attribute the steady increase in maximum visibility between 100 and 500 RC peers and after 500 RC peers to partial and full BGP feeds \cite{giotsas2014inferring}, respectively.

Analyzing the minimum visibility of MOAS prefixes shows that for around 40\% of IPv4 and  20\% of IPv6 MOAS prefixes, at least one PO pair is visible only at a single route collector peer.
For over half of IPv4 MOAS prefixes, at least one PO pair has a minimum visibility of fewer than five RC peers, and more than 90\% are visible on fewer than 100 RC peers.
Around 70\% of IPv6 MOAS prefixes have at least one PO pair with a minimum visibility of fewer than 100 RC peers.

This indicates, that each MOAS prefix is highly visible from one origin AS and has a very low visibility from another origin AS.
To confirm this, we calculate the difference between the maximum and minimum visibility \emph{for each MOAS prefix}.
As can be seen in \Cref{long:fig:max_min_visibility}, the difference distribution is very close to the maximum visibility, underlining that indeed for each MOAS prefix one origin is very visible whereas the other is barely visible.

\takeaway{Each MOAS prefix has one origin, with which the prefix is very well visible at 100 or more RC peers and another origin which is barely visible at all.
This hints at MOAS not being mainly used for anycast purposes, as for these we would expect a more balanced visibility distribution.}


\section{Users and Usage of MOAS Prefixes}
\label{sec:users-and-usage-of-moas-prefixes}

In this section we present an analysis on the users and usage of MOAS prefixes in the Internet.
Specifically, we first examine the BGP relationships between the origin ASes of MOAS prefixes.
Furthermore, we investigate the business relationships among MOAS prefix users and look into the use of MOAS prefixes by Hypergiants.
Finally, we evaluate to which extent MOAS prefixes are being utilized as anycast services.

\subsection{BGP Relationship of MOAS Origin ASes}
\label{func:originASonly-pair}

First, we want to better understand the BGP relationship between origin ASes of MOAS prefixes.
We use quarterly snapshots of CAIDA's AS to organization \cite{caidaAS2org} and AS classification inferences \cite{caidaASrelationship} datasets to identify the inter-AS BGP relationship of MOAS origin ASes.
For simplicity, we focus on MOAS prefixes with two origins.
These prefixes make up around 90\% of all MOAS prefixes and two origin ASes simplify our analysis, as we do not have to consider ternary or higher order relationships.

We check for the following three categories of BGP relationships for MOAS prefixes.
First, if both origin ASes have the same organization name in the dataset, we call them ``Siblings''.
Second, if any customer-to-provider or provider-to-customer relationship between the origin ASes of MOAS prefixes exists, we classify the inter-origin-AS relationship as ``C2P/P2C''.
Next, we examine if both origin ASes fall into the ``Peering'' category.
Finally, if we do not observe any of the above known BGP relationships between the origin AS of a MOAS prefix, we mark the inter-origin-AS BGP relationship as a ``No Relation Detected''.

\begin{figure}[!t]
  \centering
  \includegraphics[width=\linewidth]{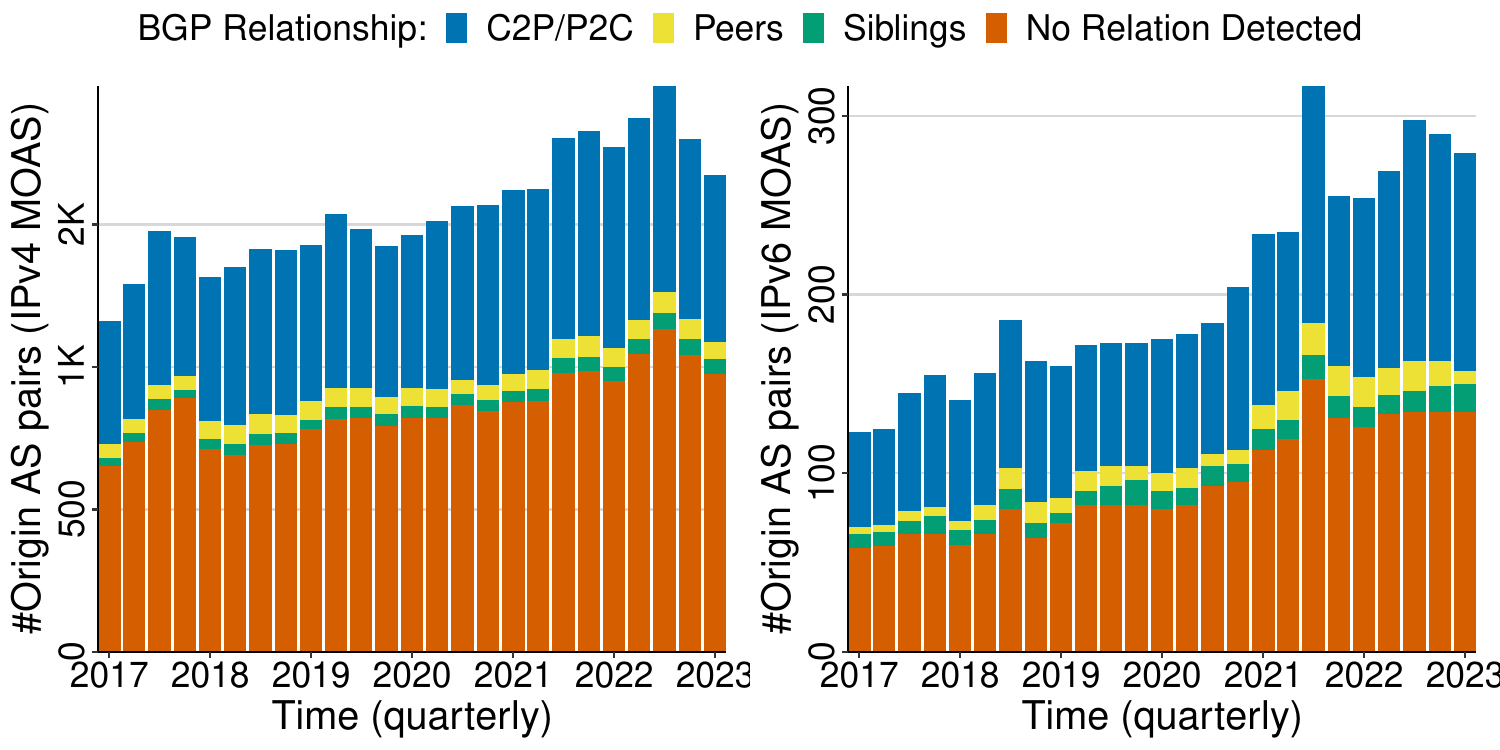}
  \caption{BGP relationship of MOAS origin AS pairs for IPv4 (left) and IPv6 (right).}
\label{fig:function:originASonly_relation}
\end{figure}

\Cref{fig:function:originASonly_relation} illustrates the inter-origin-AS BGP relationship of MOAS prefixes for IPv4 (left) and IPv6 (right) over time.
We find that almost half of all origin AS pairs for both IPv4 and IPv6 are classified as C2P/P2C (shown in blue).
Peering relationships (shown in yellow) are less frequent, with siblings making up the smallest part of MOAS origin AS pairs.
We do not detect any relationship using the CAIDA datasets for around half of all origin AS pairs.
Overall, the distribution of categories remains relatively stable over time.
Furthermore, when using the number of PO pairs as a metric (not shown), we see a similar trend  for the share of relationship categories as in \Cref{fig:function:originASonly_relation}, with a comparatively higher share of ``Siblings'' for two snapshots around 2020.


\takeaway{The vast majority of detected MOAS origin AS relationships fall into the customer to provider category.
This shows, that many MOAS prefixes are not related to sibling ASes, rather a provider announces the same prefix as the customer.}

\subsection{Relationship and Business Type of MOAS Origin ASes}
\label{subsec:relationship-and-business-type-of-moas-users}

To shed more light on the business relationship between MOAS origin AS pairs, we now apply the 17 top-level (``layer 1'') categories of the ASdb dataset \cite{ziv2021asdb} on the most recent snapshot of MOAS prefix origin ASes (January 1, 2023).
We also check MOAS origins in all available ASdb datasets for the past two years and observe a similar pattern.

As in the previous analysis, we focus on MOAS prefixes with exactly two origin ASes, \ie 2618 unique origin ASes in total.
For 97.2\%, we find a matching business category in ASdb, out of which 77.6\% match a single, and 23.4\% origin ASes match more than one category.
For easier understanding, we focus our analysis on ASes with one business category.

\begin{figure}[!t]
  \centering
  \includegraphics[width=\linewidth]{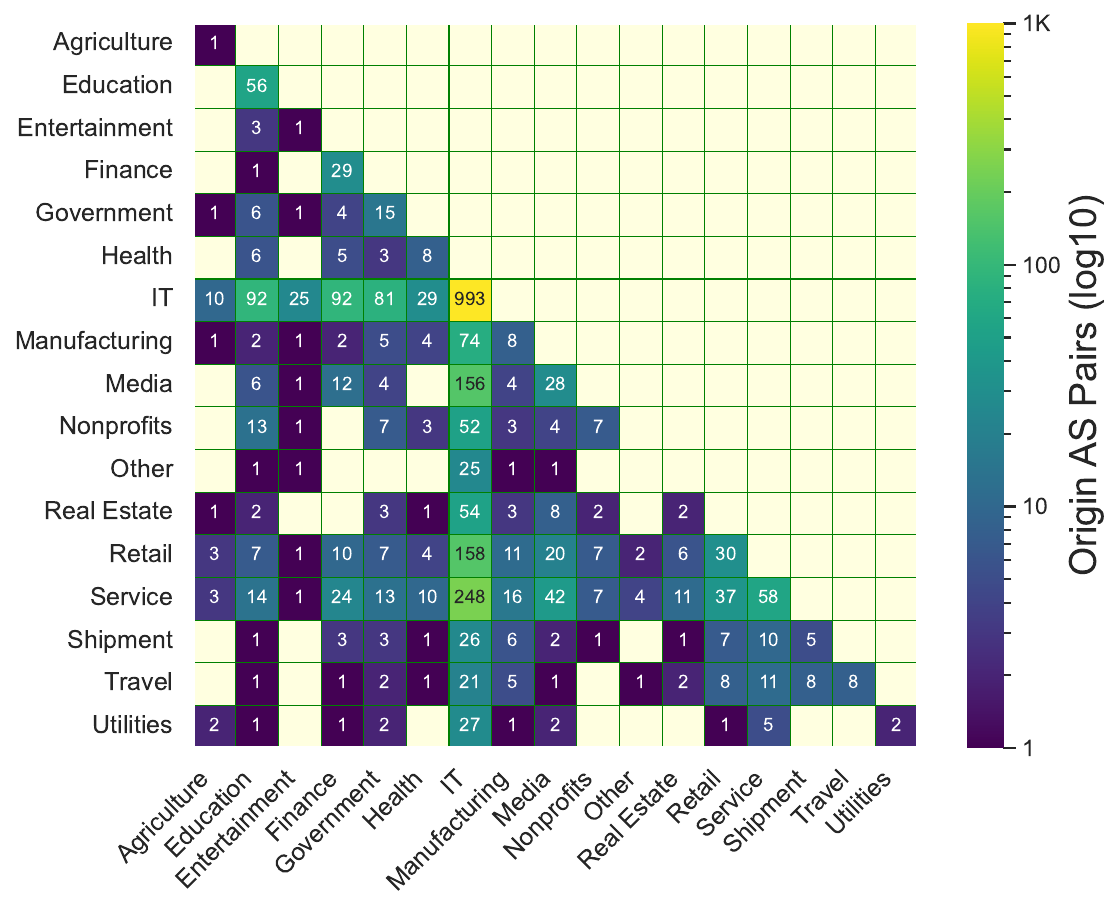}
  \caption{Heatmap of business type relationships between MOAS origin pairs.}
\label{fig:function:moas_business_relationship_origins}
\end{figure}

\Cref{fig:function:moas_business_relationship_origins} shows a heatmap of prevalent business categories for MOAS origin AS pairs.
The number and color of every cell within the heatmap represents the prevalence of each category pair for MOAS origins.
Examining the heatmap cells, we notice that in roughly 40\% of the cases, both MOAS origins fall into the ``IT'' category.
Furthermore, we notice a high number of IT company pairs with other business types, such as ``Service'', ``Retail'', or ``Media''.
This indicates that in addition to advertising their own prefix, organizations leverage IT companies to re-advertise prefixes, resulting in a MOAS event.
Moreover, for every business type, we see an elevated level on the diagonal line, which means that ASes originate MOAS prefixes from the same business type.
Overall, 89\% of MOAS prefixes have at least one origin AS classified as an IT company.
To exclude a possible bias towards IT companies, we analyze the prevalence of the IT category in the ASdb dataset.
We find that MOAS origins from the IT category have a 10 percentage points higher share than general origin ASes in the ASdb dataset (56\% vs. 46\%).

\takeaway{MOAS prefixes are mostly originated by IT companies, with the same company type for both origins being most common.}



\subsection{Hypergiants Using MOAS Prefixes}

Next, we investigate to what degree Hypergiants make use of MOAS prefixes.
Hypergiants are large content providers, CDNs, and cloud networks (\eg Google, Amazon, Meta) with a heavy-outbound traffic profile \cite{labovitz2010internet}.
We use the list of 16 Hypergiants by Gigis et al. \cite{gigis2021seven} and find that 11 of them (\ie Akamai, Alibaba, Amazon, CDNetworks, Cloudflare, Facebook, Google,  Incapsula, Limelight, Netflix, and Verizon) announce a total of 313 MOAS prefixes.
The top three Hypergiants making use of MOAS prefixes are Verizon (98 MOAS prefixes), Netflix (48), and Google  (45).
This shows that MOAS prefixes are used by big players in the Internet, possibly to improve their network's resilience, performance, and quality of experience for their users.

\takeaway{11 out of 16 Hypergiants use MOAS prefixes to reap the benefits of their users being able to access their networks via multiple origins.}


\subsection{Anycast MOAS Prefixes}
\label{subsec:anycasted-moas-prefixes}

MOAS prefixes can be good candidates for anycast services as they originate from multiple ASes. However, we find that only 225 (0.9\%) of IPv4 and 89 (6.3\%) of IPv6 MOAS prefixes are identified as anycast prefixes \cite{anycasttool}.

We observe that the number of origins per prefix for anycast MOAS prefixes is higher than for general MOAS prefixes:
While around 90\% of MOAS prefixes have two origin ASes, this percentage is less than 40\% for anycast MOAS prefixes.
More than 50\% of anycast MOAS prefixes are announced by more than ten origin ASes.
The prefix with the most origin ASes (61 origins) is the IPv6 prefix \texttt{2001:503:231d::/48}, which is attributed to Verisign.

Even though the official root-server website lists single origin ASes for each of the 13 root DNS servers \cite{rootservers}, we observe that the IPv4 address of A and J root DNS servers, operated by Verisign, use MOAS prefixes with a /24 CIDR size.
In contrast to IPv4, we do not find any IPv6 MOAS prefix for DNS root servers.

\takeaway{The use of anycast within MOAS prefixes is low, with under 1\% for IPv4 and around 6\% for IPv6.
When MOAS prefixes are used for anycast, more than 50\% of them have more than ten origins.
This shows, that MOAS prefixes are rarely used for anycast purposes, but if they are, they use a large number of origin ASes.}



\section{Related Work}
\label{sec:relatedwork}

In this section we present related work in the areas of prefix hijacking and MOAS prefixes.

\parx{Prefix Hijacking:}
When an attacker announces someone else's prefix in BGP from its own AS, the prefix is being hijacked.
A prefix hijack can cause service interruptions and consequently lead to traffic, financial, and reputation loss for the victim.
The first well-known prefix hijacking case happened in 2008 when Pakistan Telecom hijacked the YouTube prefix for censorship purposes \cite{pakistan-youtube}.
Since then multiple prefix hijacking events have occurred.
A more recent example of such an incident happened in August 2022 when several crypto services hosted on AWS were hijacked, resulting in financial loss \cite{aws-crypto-hijack}.
Previous work has suggested various techniques and tools to detect and mitigate BGP prefix hijacks \cite{sermpezis2018artemis,qin2022themis,imaiposter,testart2019profiling,oliver2022stop,biersack2012visual}.
Websites such as BGPmon \cite{bgpmon} and CAIDA's Hijack Observatory \cite{caida_hijacks_observatory} are monitoring the BGP for possible prefix hijack events.
In 2018, Sermpezis et al. conducted a survey among 75 network operators which showed that operators mostly rely on third parties such as BGPmon \cite{bgpmon} to detect prefix hijacks \cite{sermpezis2018survey}.
In 2019, Cho et al. presented a classification of BGP hijacks into different categories and found that 4\%, 1\%, and 2\% of hijacks are due to typos, prepend mistakes, and BGP hijacking with a forged AS path, respectively \cite{cho2019bgp}.

\parx{MOAS Prefixes:}
In 2001, Zhao et al. investigated MOAS prefixes for the first time on the Internet using Routeviews route collectors \cite{routeviews2022data} finding that the majority of MOAS conflicts were short-lived, lasting only a few days \cite{zhao2001analysis}.
In 2007, Chin analyzed MOAS prefixes observed over 21 days and concludes that MOAS constitute a small but growing percentage of reachability information \cite{chin2007characteristics}.
In 2011, Bornhauser et al. present MOAS Analyzer, a tool to identify, analyze, and automatically classify MOAS conflicts \cite{bornhauser2011automatic}.
In 2014, Jacquemart et al. performed a longitudinal analysis of MOAS prefixes and showed that short-lived MOAS events are not due to misconfigurations, but because of origin instability or route flapping \cite{jacquemart2014longitudinal}.
In 2015, Schlamp et al. investigated subprefix MOAS events using IRR data, topology data, and TLS certificates and found that  the majority of these announcements with multiple origins are harmless \cite{schlamp2015investigating}.

To the best of our knowledge, this is the first work to perform an in-depth longitudinal analysis of long-lived MOAS prefixes.

\section{Conclusion}
\label{sec:conclusion}

In this paper we analyzed long-lived MOAS prefixes over a period of six years.
In IPv4 we identified around 24k long-lived MOAS prefixes compared to 1.4k in IPv6.
We found that the vast majority of MOAS prefixes with entries in RPKI have a valid ROV state for all origins.
Moreover, we found MOAS prefixes to use very specific CIDR sizes, indicating their use for fine-grained traffic steering.
We identified that a considerable number of MOAS prefixes are related to mergers and acquisitions of companies.
We recommend that network operators should clean up the extra MOAS produced due to mergers or other situations to prevent unnecessarily routing entries in the Internet routing environment.
Furthermore, over 90\% of MOAS prefixes were announced by two origin ASes, with the majority of origin pairs being in a customer-provider relationship.
Finally, we have shown that the majority of MOAS users are IT companies and that MOAS are rarely used for anycast purposes.
\twemoji{vulcan_salute}


\printbibliography

\end{document}